\documentclass[twocolumn, amssymb, nobibnotes, aps, superscriptaddress,prl]{revtex4-1}

\usepackage[colorlinks=true,linkcolor=blue,citecolor=blue,urlcolor=blue]{hyperref}
\usepackage{graphicx}
\usepackage{gensymb}
\usepackage[normalem]{ulem}
\usepackage{color}
\usepackage{multirow}
\usepackage{amsmath}
\usepackage{amsfonts}
\usepackage{natbib}
\usepackage{bm}
\usepackage{textcomp}
\usepackage{multirow}
\usepackage{xcolor}
\usepackage{balance}
\usepackage{orcidlink}

\makeatletter
\newcommand*{\ClearPage}{%
  \close@column@grid
  \clearpage
  \twocolumngrid
}
\makeatother

\def\qe{\textsc{Quantum ESPRESSO}\texttrademark}

\usepackage[columnwise]{lineno}

\begin{document}

\title{
  First-principles study of the gap in the spin excitation spectrum \\ of the CrI$_3$ honeycomb ferromagnet
}

\author{Tommaso Gorni\,\orcidlink{0000-0002-7139-8429}}
\email [Corresponding author: ]{t.gorni@cineca.it}
\altaffiliation{Present address: CINECA National Supercomputing Center, Casalecchio di Reno, I-40033 Bologna, Italy, European Union}
\affiliation{LPEM, ESPCI Paris, PSL Research University, CNRS, Sorbonne Universit\'e, 75005 Paris France, European Union}
\author{Oscar Baseggio\,\orcidlink{0000-0002-5733-2001}}
\affiliation{SISSA -- Scuola Internazionale Superiore di Studi Avanzati, Trieste, Italy, European Union}
\author{Pietro Delugas\,\orcidlink{0000-0003-0892-7655}}
\affiliation{SISSA -- Scuola Internazionale Superiore di Studi Avanzati, Trieste, Italy, European Union}
\author{Iurii Timrov\,\orcidlink{0000-0002-6531-9966}}
\affiliation{{T}heory and Simulation of Materials (THEOS) and National Centre for Computational Design and Discovery of Novel Materials (MARVEL), \'{E}cole Polytechnique F\'{e}d\'{e}rale de Lausanne, CH-1015 Lausanne, Switzerland}
\author{Stefano Baroni\,\orcidlink{0000-0002-3508-6663}}
\email{baroni@sissa.it}
\affiliation{SISSA -- Scuola Internazionale Superiore di Studi Avanzati, Trieste, Italy, European Union}
\affiliation{CNR -- Istituto dell'Officina dei Materiali, SISSA, Trieste, Italy, European Union}

\begin{abstract}
  The nature of the gap observed at the zone border in the spin-excitation spectrum of CrI$_3$ quasi-2D single crystals is still controversial. We perform first-principles calculations based on time-dependent density-functional perturbation theory, which indicate that the observed gap results from a combination of spin-orbit and inter-layer interaction effects. The former give rise to the anisotropic spin-spin interactions that are responsible for its very existence, while the latter determine both its displacement from the K point of the Brillouin zone  due to the in-plane lattice distortions induced by them, and an enhancement of its magnitude, in agreement with experiments and previous theoretical work based on a lattice model.
\end{abstract}

\date{\today}

\maketitle

The recent discovery of two-dimensional (2D) magnets~\cite{Huang:2017, Gong:2017} has opened new paths in nano\-device fabrication and engineering, thanks to the high tunability of their magnetic state and the ease of their incorporation into functional heterostructures~\cite{Gibertini:2019, Burch:2018, Novoselov:2016}. In these materials, the magnetic anisotropy induced by spin-orbit couplings (SOC) stabilizes the 2D long-range magnetic order against thermal fluctuations, by gapping the magnon Goldstone mode and thus dodging the conclusions of the Mermin-Wagner theorem~\cite{Mermin:1966,Hohenberg:1967}. The van der Waals crystal CrI$_3$ was the first compound reported to display long-range magnetic order down to the monolayer (ML) limit, where it behaves as a ferromagnetic semiconductor with a Curie temperature of $45$\,K and a sizeable out-of-plane anisotropy~\cite{Huang:2017}. 

Fascinating prospects arise from the potential of CrI$_3$ to sustain collective magnetic excitations, namely magnons. Inelastic neutron scattering (INS) measurements on rhombohedral quasi-2D CrI$_\mathrm{3}$ single crystals~\cite{Chen:2018,Chen:2021} show that magnons in CrI$_3$ present a two-band spin-wave dispersion, whose symmetries confirm the  expected picture of CrI$_3$ as a 2D honeycomb Heisenberg lattice formed by Cr local moments. Very intriguingly, the band crossings at the zone-corners---the K points of the 2D Brillouin zone (BZ)---predicted by the simplest model featuring only nearest-neighbour in-plane exchange interactions~\cite{Fransson:2016} are not observed. The measurements find instead a finite gap---first reported as $\sim 4 \ \mathrm{meV}$~\cite{Chen:2018} and then refined to $2.8 \ \mathrm{meV}$~\cite{Chen:2021}---between the two branches. 
Whether this feature subsists for CrI$_3$ ML, as expected for weak inter-layer magnetic interactions, has not yet found experimental confirmation, due to the inefficacy of INS in the ML regime.
Moreover, it was recently remarked that the measured gap size is extremely sensitive to experimental conditions including sample mosaic, resolution, and momentum integration range~\cite{Do:2022}.

The microscopic origin of the gap observed at the bulk's BZ zone corners---as well as its very existence in the ML limit---is still controversial and under active theoretical scrutiny~\cite{Soriano:2020}. Much of the work performed so far to settle this issue is based on the Heisenberg model for the 2D honeycomb lattice, with anisotropic exchange interactions of various forms, ranging from Dzyalonshinskii-Moriya (DM)~\cite{Kim:2016,Owerre:2016,Chen:2018,Chen:2021} to  Kitaev and anisotropic symmetric exchange~\cite{Xu:2018,Olsen:2019,Lado:2017,Lee:2020,Aguilera:2020}. Remarkably, some of these models predict a splitting between the otherwise degenerate acoustic and optical magnon modes at the K point of the 2D BZ, giving rise to topologically protected edge excitations within the gap, in analogy with the Haldane model for electrons~\cite{Haldane:1988,Kane2005}. Also, it has been shown that inter-layer couplings may have a significant effect on the magnon dispersion in the bulk---where spin-wave excitations have only been detected so far---by shifting the band crossings/edges off the 2D BZ zone corners~\cite{Ke:2021}.

The use of density-functional theory (DFT) for estimating the exchange couplings in these models may yield quite variable results, depending on the model assumptions and on the functional being adopted~\cite{Lado:2017,Xu:2018,Olsen:2019,Soriano:2020,Besbes:2019, Kvashnin:2020}. In general, standard exchange-correlation (XC) functionals based on local spin-density approximation (LSDA) and generalized-gradient approximation (GGA) yield too large a spin stiffness~\cite{Yin:2011,Singh:2019}, which in the case of CrI$_3$ can be significantly redressed by on-site Hubbard corrections for Cr $3d$ {states}~\cite{Ke:2021,Olsen:2019}; for what concerns the anisotropic exchange, the most recent calculations find very weak~\cite{Kvashnin:2020} if not negligible~\cite{Pizzochero:2020} DM/Kitaev interactions. Other studies avoid the use of lattice models and are based instead on time-dependent DFT (TDDFT)~\cite{Ke:2021} or many-body perturbation theory (MBPT)~\cite{Olsen:2021}. The  TDDFT study of bulk CrI$_3$ in Ref.~\cite{Ke:2021}, performed by neglecting SOC, predicts that, while no global gap is observed at this level of theory, a proper account of inter-layer couplings shifts the band crossings away from the BZ corners, which would thus display a local gap. In contrast, a study based on a solution of the Bethe-Salpeter equation including SOC for a CrI$_3$ ML (thus no inter-layer coupling) predicted a gap of 0.3~meV at the K point of the BZ~\cite{Olsen:2021}. In spite of all these efforts, the current understanding of the phenomenology of the spin dynamics in this intriguing system relies on a combination of semiempirical models and \emph{ab initio} calculations that, not accounting for SOC effects in the bulk, miss the crucial element necessary to explain the experimental data.

\begin{figure}[t]
    \begin{center}
        \includegraphics[width=0.5\textwidth]{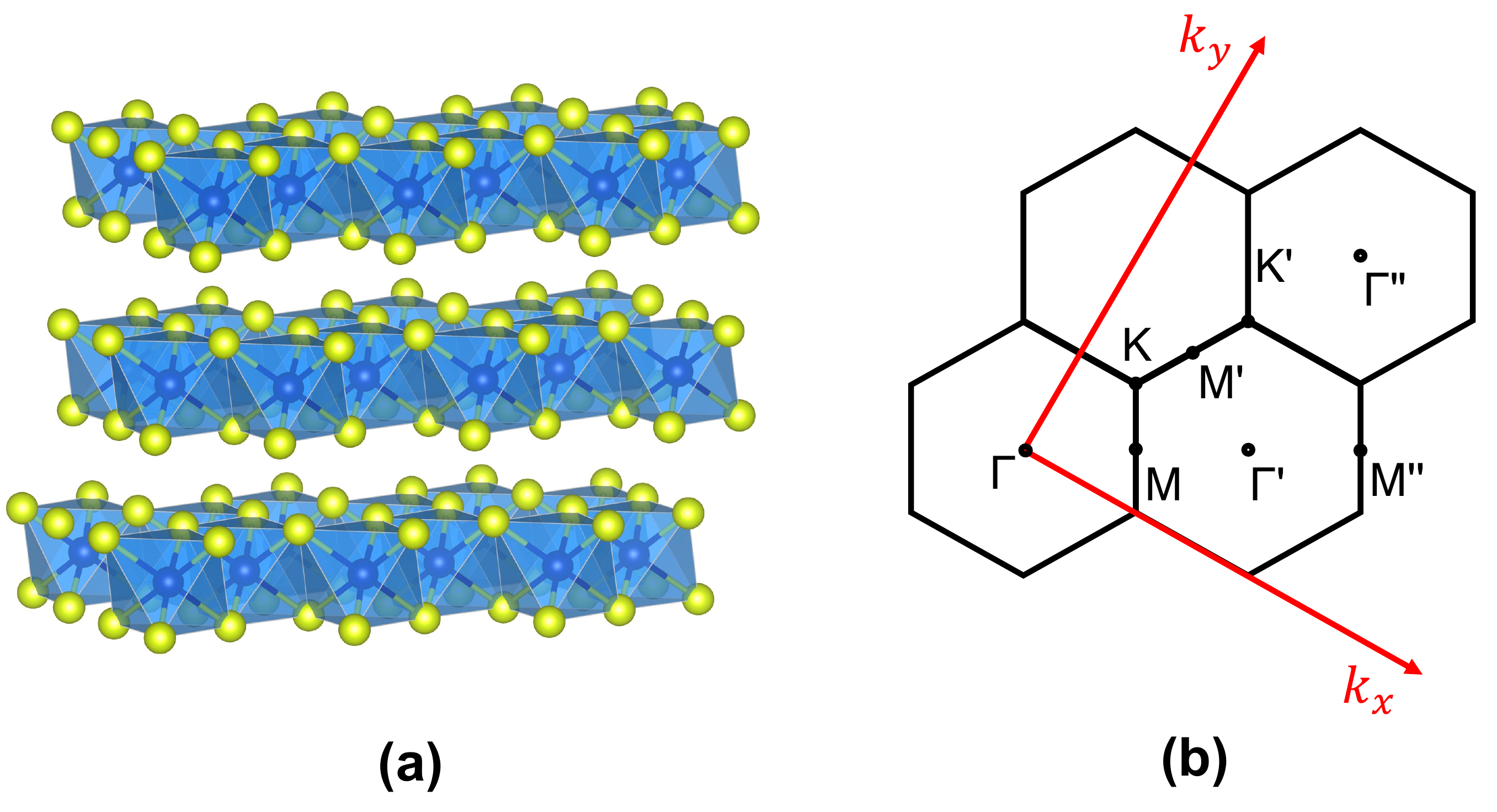}
        \caption{(a)~Crystal structure of the bulk CrI$_3$, where Cr and I atoms are shown in blue and yellow. (b)~2D Brillouin zone of the hexagonal cell. High-symmetry points are also shown together with the Cartesian framework.
        }
        \label{fig:structure}
    \end{center}
\end{figure}

In this letter we address the microscopic origin of the gap observed in the spin-wave excitation spectrum of bulk CrI$_3$ by using a novel implementation of TDDFT~\cite{Gorni:2016,Gorni:2018}, based on the Liouville-Lanczos technique~\cite{Walker:2006,Rocca:2008}. Our approach is fully \emph{ab initio} and accounts for leading-order relativistic corrections without relying on any adiabatic spin decoupling, thus avoiding the intricacies of downfolding to an effective spin model and accounting for the complexity of SOC-induced exchange couplings directly into the excitation spectra, without introducing any semi-empirical parameters. All the ground- and excited-state calculations have been performed using the \qe\ (QE) distribution of computer codes~\cite{Giannozzi:2009, Giannozzi:2017, Giannozzi:2020}. We generated fully relativistic norm-conserving pseudopotentials with the \texttt{atomic} QE code, using the atomic configurations from the \texttt{PSlibrary} v0.3.1~\cite{DalCorso:2014}. Kohn-Sham spinor wavefunctions and spin and charge densities were expanded in plane waves with a kinetic-energy cutoff of 60 and 240~Ry, respectively (see Sec.~S1 in the supplemental material (SM)~\cite{Note_SM} for the convergence tests). The low-temperature, ferromagnetic phase of bulk CrI$_3$ has a rhombohedral crystal structure, leading to an ABC stacking of the Cr honeycomb sublattices (see Fig.~\ref{fig:structure})~\cite{Soriano:2020}. We used a uniform $\Gamma$-centered $8 \times 8 \times 8$ $\bm{k}$-point mesh to sample the BZ. The dimensions of the unit cell---determining the inter-layer distance and the in-plane lattice parameter---have been fixed to the experimental values \cite{McGuire:2015}. Internal coordinates have been further optimized with the \texttt{PW} QE code using the spin-polarized GGA with the Perdew-Burke-Ernzerhof (PBE) XC functional~\cite{Perdew:1996}.  These calculations correctly predict an anisotropic ferromagnetic (FM) ordering with the easy axis perpendicular to the atomic planes. The magnon spectra have been computed with the \texttt{turboMagnon} QE code \cite{Gorni:2022}, using the LSDA XC functional \cite{Perdew:1981}, and broadened by a Lorentzian smearing function with a full width at half maximum of 0.3 meV. The magnon dispersion along the high-symmetry directions was obtained by computing magnon energies on a finite set of transferred momenta and then interpolated into a finer grid (see Sec.~S3 in the SM). The sensitivity of the magnetic spectra to the in-plane lattice parameter is discussed in Sec.~S2 in the SM. The data used to produce the results of this work are available in the Materials Cloud Archive~\cite{MaterialsCloudArchive2023v2}.

\begin{figure}[t]
    \begin{center}
        \includegraphics[width=0.4\textwidth]{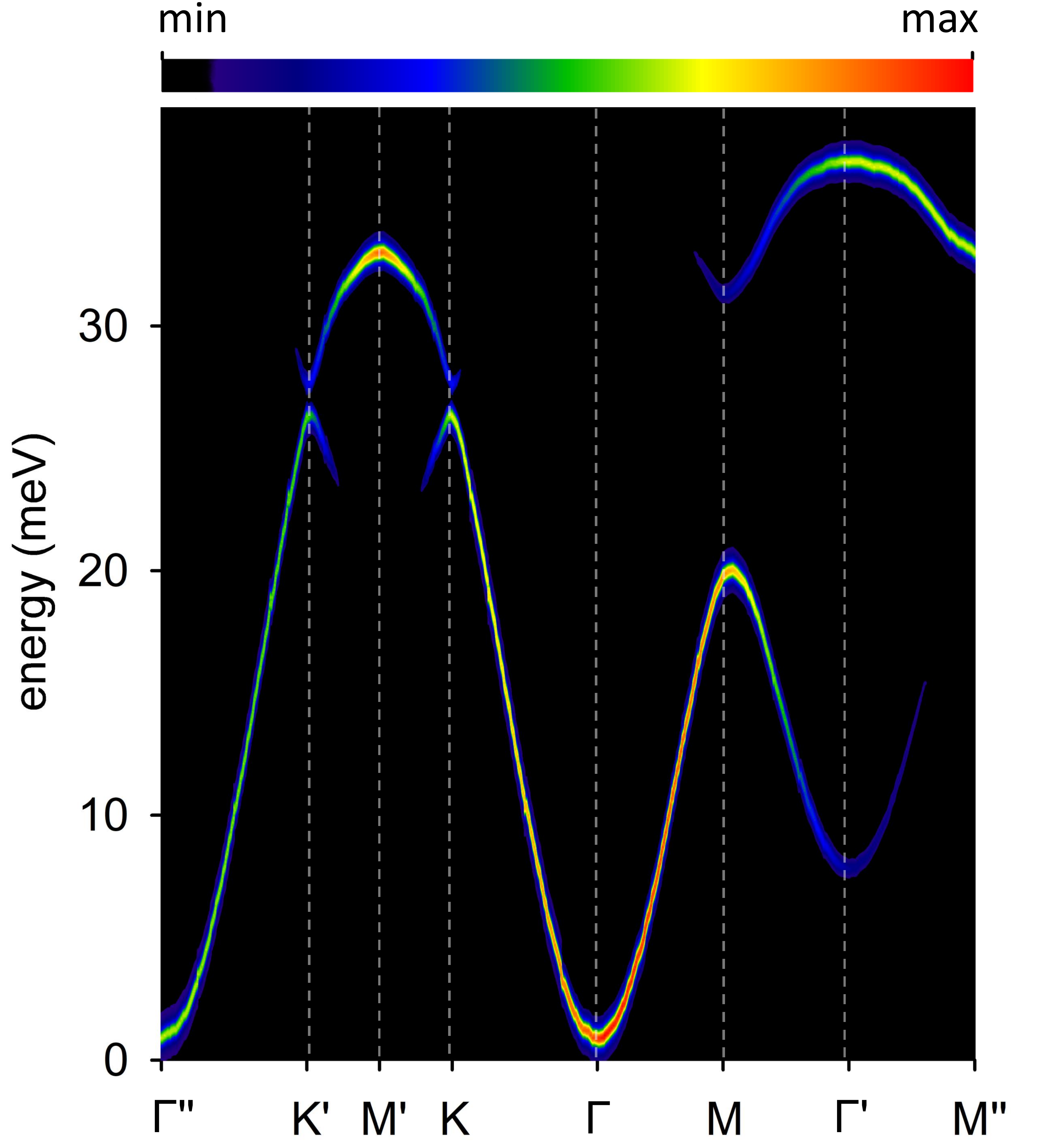}
        \caption{Magnon dispersions along the  high-symmetry directions of  the (hexagonal) BZ computed using TDDFT including SOC in the rhombohedral ferromagnetic CrI$_3$. The intensity color map is shown at the top.
        }
        \label{fig:TDDFpT_spectrum}
    \end{center}
\end{figure}

\begin{figure}[t]
 \begin{center}
    \includegraphics[width=0.4\textwidth]{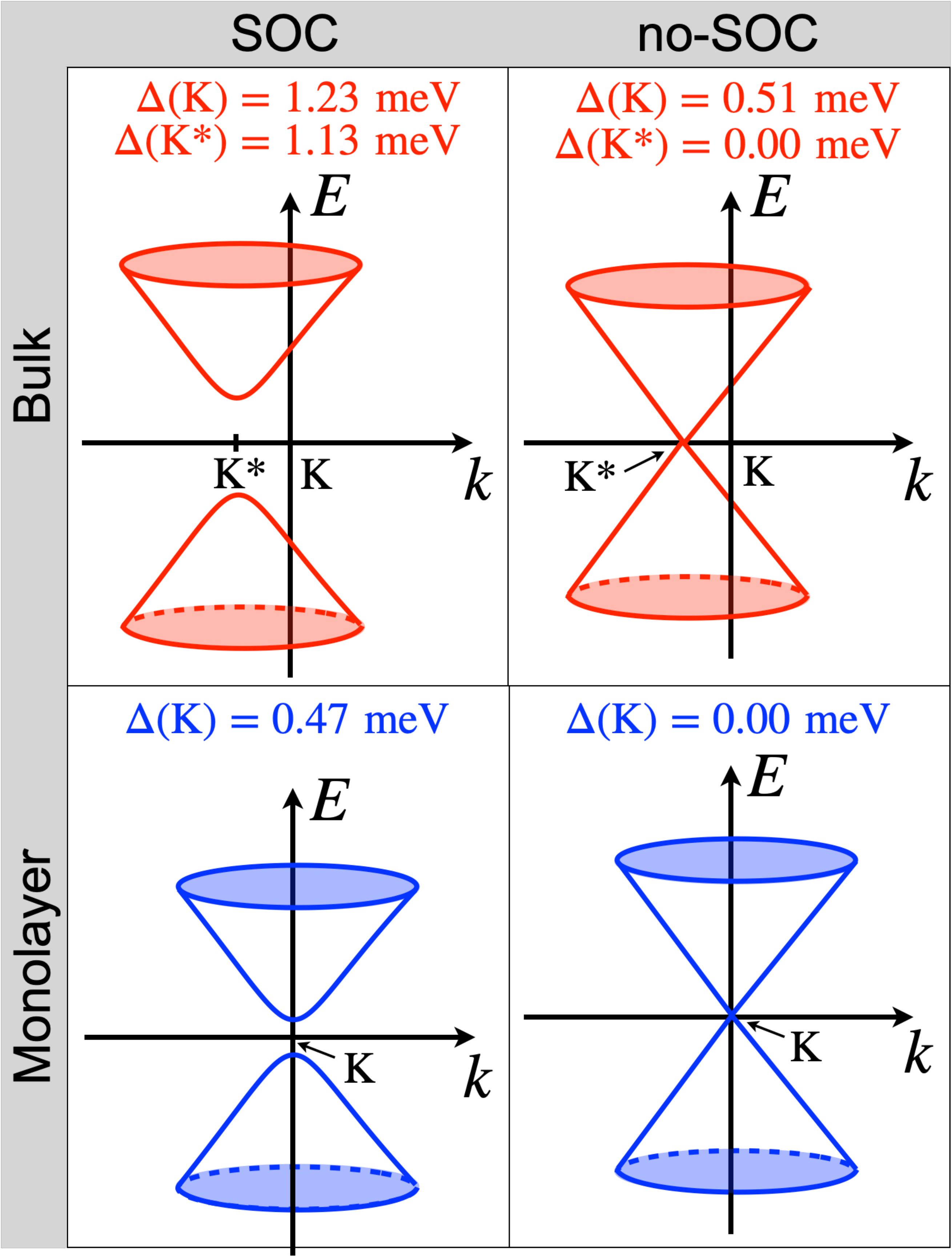}
    \caption{Schematic illustration of the magnon dispersion in bulk and ML CrI$_3$ with and without spin-orbit coupling (SOC). The high-symmetry points K and K$^*$ are highlighted. (a)~Bulk with SOC, (b)~ML with SOC, (c)~bulk without SOC, and (d)~ML without SOC. The values of gaps at K and K$^*$ are also highlighted as $\Delta(\mathrm{K})$ and $\Delta(\mathrm{K}^*)$, respectively.} 
    \label{fig:schematic_figure}
 \end{center}
\end{figure}

Our computations of the magnon spectrum in bulk CrI$_3$ yield the spin-wave dispersions reported in Fig.~\ref{fig:TDDFpT_spectrum}. 
 Following the convention adopted in previous works (e.g. Refs.~\cite{Chen:2018,Chen:2021,Ke:2021}), we plot the magnon dispersion using the high-symmetry points and directions of the hexagonal BZ of the ML, as shown in Fig.~\ref{fig:structure}. We note that symmetry points which would be equivalent in a truly honeycomb 2D structure may not be equivalent in our bulk spectra (e.g.\ $\Gamma$ and $\Gamma'$), as long as inter-layer interactions are sufficiently strong.
The global minimum of the magnon dispersion corresponds to the Goldstone mode at $\Gamma$, which has an energy of $\sim \mathrm{1.3~meV}$~\cite{Gorni:2022}. Such a finite value has to be fully ascribed to the SOC anisotropy, since when computing the same limit without SOC we obtain instead a vanishing energy, as expected for isotropic ferromagnets~\cite{Gorni:2022}. No Dirac cones are observed at the K point of the BZ, which instead features a gap of $\Delta(\mathrm{K})=\mathrm{1.23~meV}$ with a midgap energy of $\sim \mathrm{27~meV}$ (see Fig.~\ref{fig:schematic_figure}(a)). In a recent paper~\cite{Ke:2021}, Ke and Katselson suggested that, when SOC are neglected, the gap may be only apparent, for the Dirac point may simply be shifted off the K point towards a low-symmetry position, as a consequence of the slight break of the in-plane honeycomb symmetry induced by inter-layer exchange coupling. In order to ascertain if this is the case, and to find the exact location of the band edge/crossing, we have sampled the TDDFT magnon dispersion in the proximity of the K point, and fitted the resulting energies to a double conoid:
\begin{equation}
    \label{eq:conoid}
    \omega(\bm{k}) = \pm \sqrt{\alpha + (\bm{k} -\bm{k}_{\mathrm{K}^*})\cdot \bm{Q} \cdot(\bm{k}-\bm{k}_{\mathrm{K}^*})},
\end{equation}
where $\bm{k}$ is a point in the 2D BZ, $\bm{Q}$ is a symmetric $2\times 2$ tensor, $\bm{k}_{\mathrm{K}^*}$ the position of the band edge/crossing, and $\alpha$ a positive constant. $\mathrm{K}^*$ is a Dirac point when $\alpha=0$, a band edge otherwise. We find that, when SOC are neglected, the gap $\Delta(\mathrm{K}) = 0.51~\mathrm{meV}$ is apparent and we have indeed $\Delta(\mathrm{K}^*)=0$, with $\bm{k}_{\mathrm{K}^*}=\bm{k}_\mathrm{K}
+(-0.001,0.008) \frac{2\pi}{a}$, 
where $\bm{k}_\mathrm{K}=(\frac{1}{3},\frac{1}{\sqrt{3}})\frac{2\pi}{a}$ are the coordinates of the K point, and $a=6.87~\mathrm{\AA}$ is the lattice parameter of the 2D honeycomb lattice; when SOC are explicitly accounted for, we find $\Delta(\mathrm{K}^*)=1.13~\mathrm{meV}$ and  $\bm{k}_{\mathrm{K}^*}=\bm{k}_\mathrm{K}+(-0.002,0.009)\frac{2\pi}{a}$.
These displacements of less than $0.01~\mathrm{\AA^{-1}}$ along the $k_y$ direction are quite similar in the two cases, and confirm the predictions of Ke and 
Katsnelson~\cite{Ke:2021}. This is illustrated schematically in Fig.~\ref{fig:schematic_figure}. More details can be found in Sec.~S4 of the SM. 

These findings are in qualitative agreement with the experiments of Ref.~\cite{Chen:2018}, where the highest intensity in the constant-energy cuts of spin waves are observed slightly shifted outside K (see Fig.~2(b) in Ref.~\cite{Chen:2018}). According to the same reference, the lower and upper band edges seem to display two different $\bm{k}_{\mathrm{K}^*}$; we have not found however any related discussion in the literature and our TDDFT data do not seem to support this scenario (see Sec.~S4 in the SM).

\begin{figure}[t!]
    \begin{center}
        \includegraphics[width=0.4\textwidth]{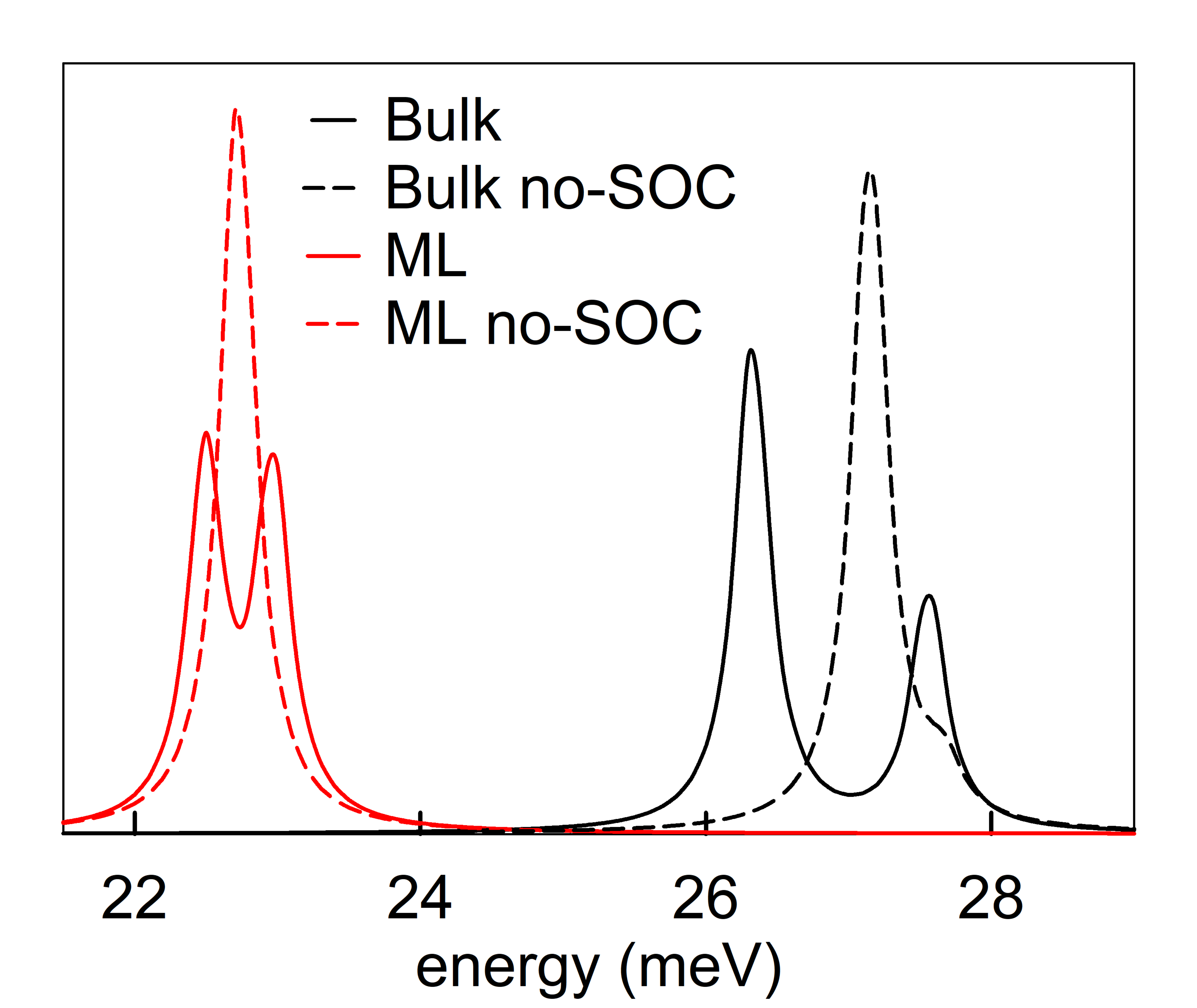}
        \caption{Magnon spectra at the high-symmetry point K computing with TDDFT in ML (red line) and bulk (in black line) CrI$_3$, solid line with spin orbit and dashed line without spin orbit.}
        \label{fig:gap_K}
    \end{center}
\end{figure}

\begin{figure}[t]
    \centering
    \includegraphics[width=0.45\textwidth]{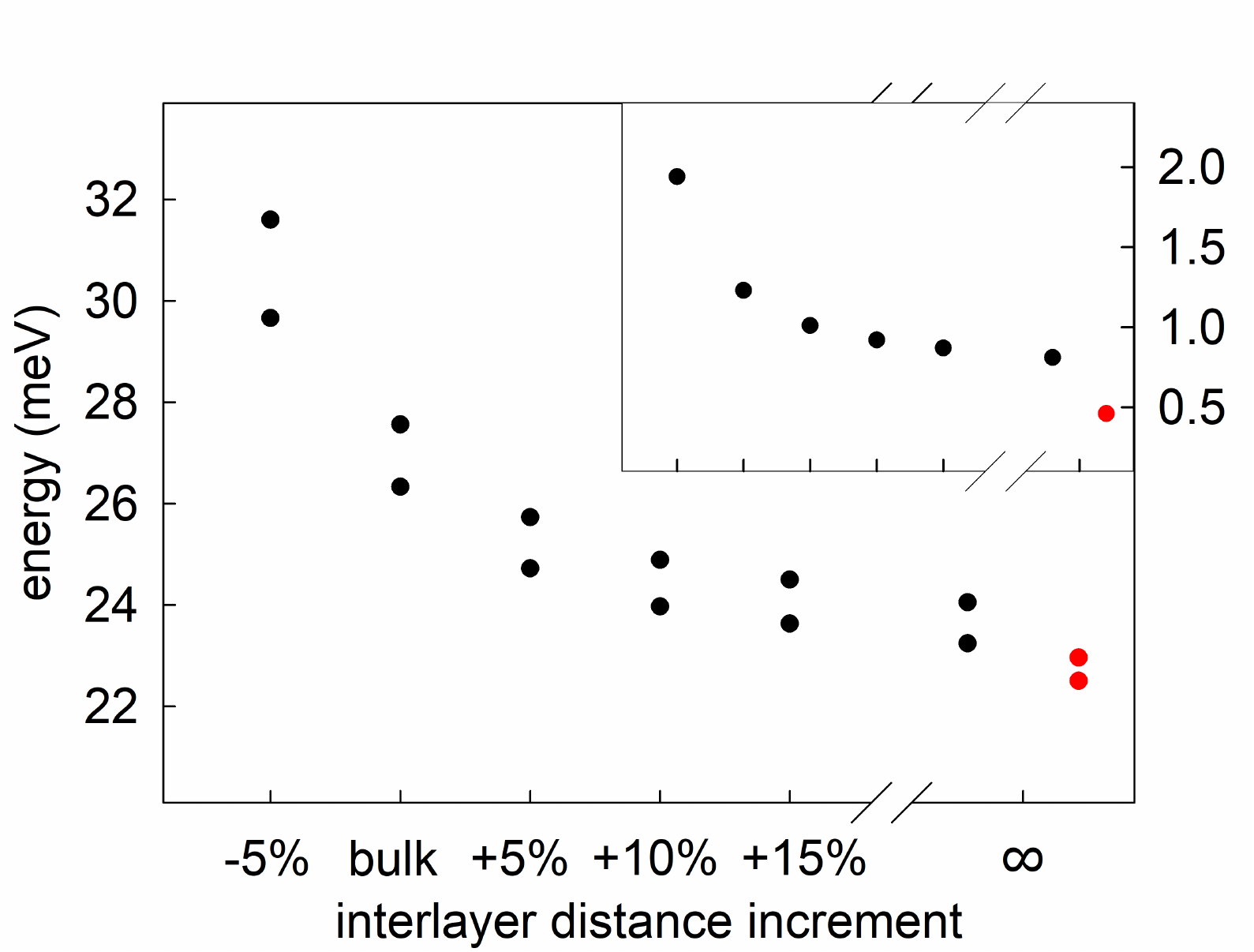}
    \caption{Magnon energies at K point as a function of the inter-layer distance. The inset shows the gap variation. Bulk data are shown in black while the fully symmetric ML is shown in red. }
    \label{fig:gap_distance}
\end{figure}

In order to establish the role of inter-layer couplings, we computed the magnon spectrum at K in a model for ML CrI$_3$ obtained by increasing the inter-layer distance to $\approx 20$~\AA, which is representative of the ML limit, and following the same methodology as in the bulk. Our results are illustrated in Fig.~\ref{fig:gap_K}. We found a threefold reduction of the magnon gap at K, down to a value of $\approx 0.47$~meV (see Fig.~\ref{fig:schematic_figure}), which points towards a significant role played by inter-layer interactions in the opening of the Dirac gap. Note the shoulder on the high-energy side of the peak of the bulk spectrum, when no SOC are considered, which is absent in the ML case. This peak is a signature of the Dirac cone that has shifted off the high-symmetry point, as illustrated in Fig. \ref{fig:schematic_figure}.  The interaction between layers can impact spin fluctuations in two distinct ways: either directly, by giving rise to inter-layer exchange couplings, or indirectly, by altering the in-plane atomic geometry, which in turn modifies the in-plane exchange couplings. In order to single out the relative importance of the two mechanisms, we computed the spin-fluctuation spectrum at K as a function of the inter-layer distance, starting from slightly below its value in the bulk, up to the ML limit, while keeping the in-plane geometry frozen to the positions that atoms would have in the bulk.
Our results are reported in Fig.~\ref{fig:gap_distance}. We observe a steady reduction of the gap, which reaches the $0.81$~meV infinite-separation limit, almost twice as large as the value computed for the fully relaxed ML, whose data are reported in red in the figure. This indicates that the rearrangement of the in-plane geometry induced by the proximity of different atomic layers increases the energy gap by almost 100\%, while inter-layer magnetic couplings both increase the magnitude of the SOC-induced gap of a comparable amount and shift it away from the K high-symmetry position. Remarkably, without SOC, no gap is observed at K in the ML (see Fig.~\ref{fig:schematic_figure}).

As expected, the major discrepancy between our results and the INS data are due to the overestimate of the spin stiffness, which produces too broad a bandwidth of $\approx 35$~meV (consistently with Ref. \cite{Ke:2021} that reports a value of $\approx 31$~meV in bulk CrI$_3$ without SOC), to be compared with the experimental value of $\approx 20$~meV~\cite{Chen:2018, Chen:2020, Chen:2021}. As previously discussed, too large a spin stiffness appears to be a common shortcoming of LSDA- and GGA-based approaches, partially redressed by a proper account of on-site Coulomb effects \cite{Ke:2021}.
Nonetheless, these effects act mainly as a rigid renormalization of the magnon dispersions, without altering the key features of the individual branches nor the nature of the magnetic coupling giving rise to them.
This discrepancy withstanding, the qualititative features of our results are in good agreement with the INS experiments of Refs.~\cite{Chen:2018,Chen:2021} and shed light onto the microscopic mechanisms responsible for the opening of the gap in the spin-fluctuation spectrum in the vicinity of the K point of the 2D BZ.

Our results can be summarized as follows: \emph{i)}~in the absence of SOC, the spin-fluctuation spectrum of ML CrI$_3$ displays no gap at the zone center and there is a Dirac cone with the bands crossing at the K point of the 2D BZ;
\emph{ii)}~SOC effects open a gap both in the Goldstone mode at the zone center and in the Dirac cone exactly at the K point of the 2D BZ in ML; 
\emph{iii)}~In the bulk without SOC, there is no gap at the zone center, there is an apparent gap exactly at K and the crossing of bands of the Dirac cone is shifted outside of K due to the inter-layer interactions;
\emph{iv)}~In the bulk, SOC effects open a gap in the Goldstone mode at the zone center and open a gap in the Dirac cone with band edges shifted outside of K due to inter-layer interactions;
\emph{v)} in-plane lattice distortions due to the proximity of neighboring atomic layers considerably enhance the widening effect. 
This last point makes us speculate that spin-lattice coupling may be sizeable in CrI$_3$ and leave additional signatures in its spin-wave spectra. Indeed, spin-lattice couplings have been advocated to interpret anomalies in the magnetic susceptibility~\cite{McGuire:2017}, Raman modes~\cite{McCreary:2020,Kozlenko:2021}, and magnon linewidths~\cite{Chen:2021b} in chromium trihalides, where also \emph{ab initio} simulations have shown a net dependence of Raman modes to the underlying magnetic order~\cite{Webster:2018}. A recent work from the authors of this paper shows by \emph{ab initio} simulations that SOC-induced spin-lattice couplings can induce $\approx$\,meV hybridization gaps in the magnon dispersion of ML CrI$_3$, suggesting that similar phenomena might be at play in the bulk form as well~\cite{Delugas:2023}.

\smallskip This work was partially funded by the European Union through the \emph{ \textsc{MaX} Centre of Excellence for Supercomputing applications} (project No. 824143), by the Italian MUR through the PRIN 2017 \emph{FERMAT} (grant No. 2017KFY7XF) and the \emph{National Centre from HPC, Big Data, and Quantum Computing} (grant No. CN00000013), and by the Swiss National Science Foundation (SNSF), through grant No. 200021-179138, and its National Centre of Competence in Research (NCCR) MARVEL.

\bibliography{biblio}

\ClearPage
\widetext

\setcounter{equation}{0}
\setcounter{figure}{0}
\setcounter{table}{0}
\setcounter{page}{1}
\setcounter{section}{0}
\makeatletter
\renewcommand{\theequation}{S\arabic{equation}}
\renewcommand{\thefigure}{S\arabic{figure}}
\renewcommand{\bibnumfmt}[1]{[S#1]}
\renewcommand{\citenumfont}[1]{S#1}
\renewcommand{\thesection}{S-\Roman{section}}
\renewcommand{\thepage}{S\arabic{page}}
\onecolumngrid

\begin{center}
\Large\textbf{
Supplemental Material for ``First-principles study of the gap in the spin excitation spectrum of the CrI$_3$ honeycomb ferromagnet''
}
\\[5pt]
\small
Tommaso Gorni, Oscar Baseggio, Pietro Delugas, Iurii Timrov, and Stefano Baroni
\end{center}

\section{Convergence of magnetic spectra with respect to the kinetic-energy cutoff}
\label{sec:cutoff}

\begin{figure}[ht]
 \centering
        \includegraphics[width=0.45\textwidth]{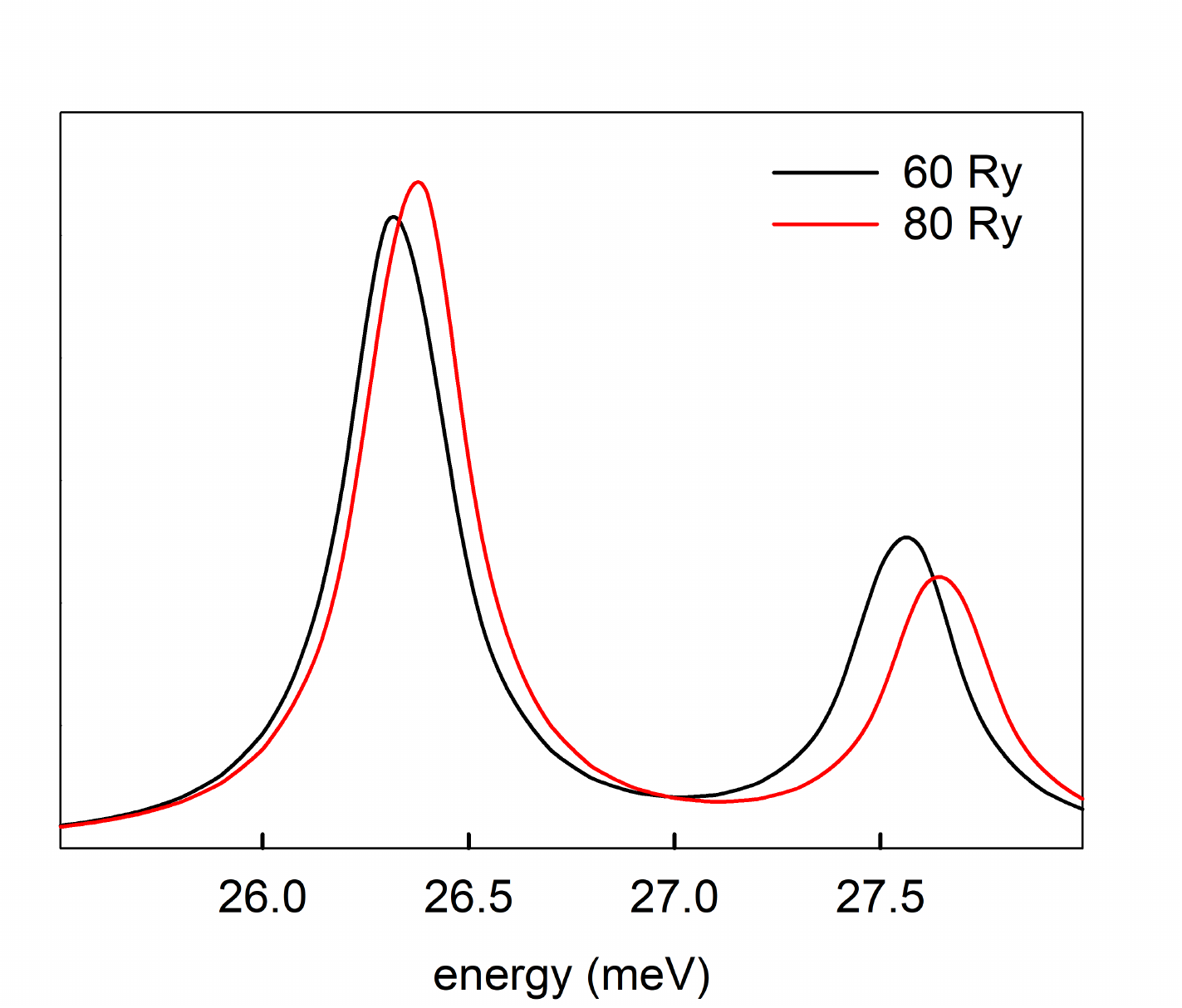}
        \caption{Comparison of the ALDA-TDDFT magnetic spectra of bulk CrI$_3$ at the high-symmetry point K computed using two values of the kinetic-energy cutoff, 60 and 80~Ry.\label{fig:conv}}
\end{figure}

Since a sub-meV resolution is needed to analyze the magnon dispersions in CrI$_3$ it is crucial to check the convergence of the magnetic spectra with respect to the kinetic-energy cutoff. In Fig.~\ref{fig:conv} we show a comparison of the magnetic spectra for bulk CrI$_3$ at the K point in the Brillouin zone computed using two values of the cutoff, 60 and 80~Ry, within the adiabatic local density approximation (ALDA).  
The magnon gap is converged within $0.05$\,meV, with the individual peak positions shifting of $0.05-0.09$\,meV at most.
For this reason, in the main text we present calculations that were all done at 60~Ry. 

\section{Sensitivity of the magnetic spectra to the in-plane lattice parameter}
\label{sec:lat_param}

\begin{figure}[ht]
    \centering
        \includegraphics[width=0.45\textwidth]{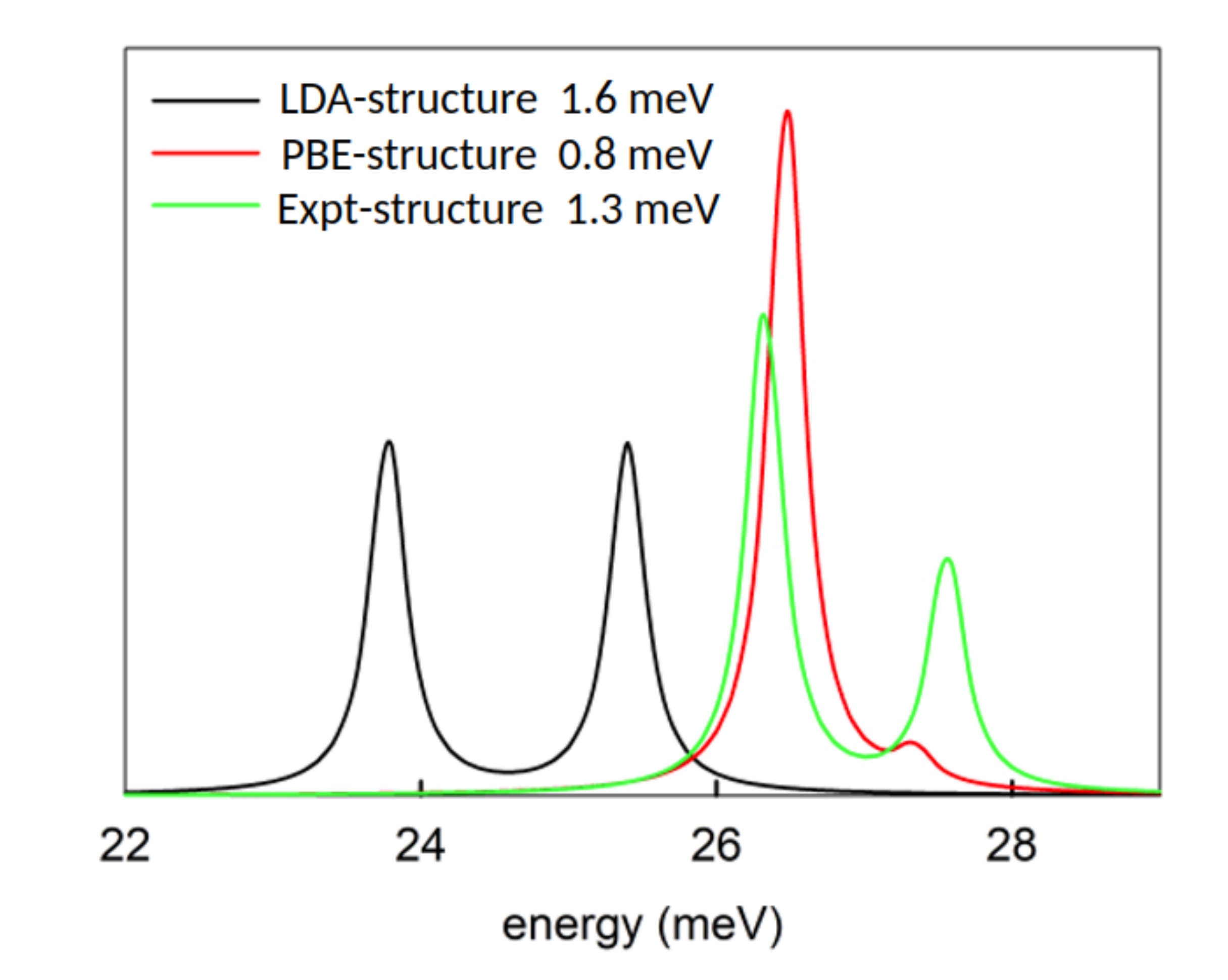}
        \caption{Comparison of the magnetic spectra of bulk CrI3 at K 
        computed at the optimized geometries corresponding to three different in-plane 
        lattice parameters. The green curve corresponds to the experimental geometry 
        used in the main text; the black corresponds to the in-plane lattice constant 
        obtained with relativistic LDA; the red curve corresponds to the PBE-optimized 
        one. We report the estimated gap values in the legend.}
        \label{fig:SM_conv}
\end{figure}

The CrI$_3$ bulk structure considered in the main text has been built by using the experimental value for the in-plane and inter-plane lattice constants, and we have then relaxed the internal atomic positions using the PBE functional. 
In order to assess the robustness of our results with respect to this choice, we report in Fig.~\ref{fig:SM_conv} the magnetic spectra for the rhombohedral CrI$_3$ with three different in-plane hexagonal lattice constants: $(i)$~the experimental one used in the main paper ($a_\mathrm{expt} = 12.9768$~Bohr), $(ii)$~the reduced one ($a_\mathrm{LDA} = 12.7556$~Bohr) obtained by energy minimization using the noncollinear relativistic LDA functional, and $(iii)$~the expanded one ($a_\mathrm{PBE} = 13.3025$~Bohr)  obtained by energy minimization using the spin-polarized PBE functional. 
As both the LDA and PBE functionals yield poor a description of the van der Waals interactions, we have chosen to keep the inter-layer distance fixed at the experimental value.
Once fixed the $a$ and $c$ parameters, the internal coordinates have been relaxed with PBE in case $(i)$ and $(iii)$, and with LDA in case $(ii)$.
The so-obtained structures have been used to compute the TDDFT magnon spectrum at the K point within the ALDA.
The results indicate that a smaller lattice constant increases the branch separation, while a larger one decreases it, suggesting a direct dependence between the in-plane lattice constant and the gap width.
Moreover, all the three cases display a magnon gap at the K point (which is an upper bound to the actual gap value, as explained in the main text) in the order of 1\,meV, showing the robustness of our claim, namely that the combination of interlayer interactions and SOC open a meV-like gap in proximity to the Dirac point in bulk CrI$_3$, with respect to the choice of the functional for the structural relaxation.

\section{Computation of the spin-wave dispersion along high symmetry directions}

\begin{figure}[ht]
    \centering
    \includegraphics[width=0.45\textwidth]{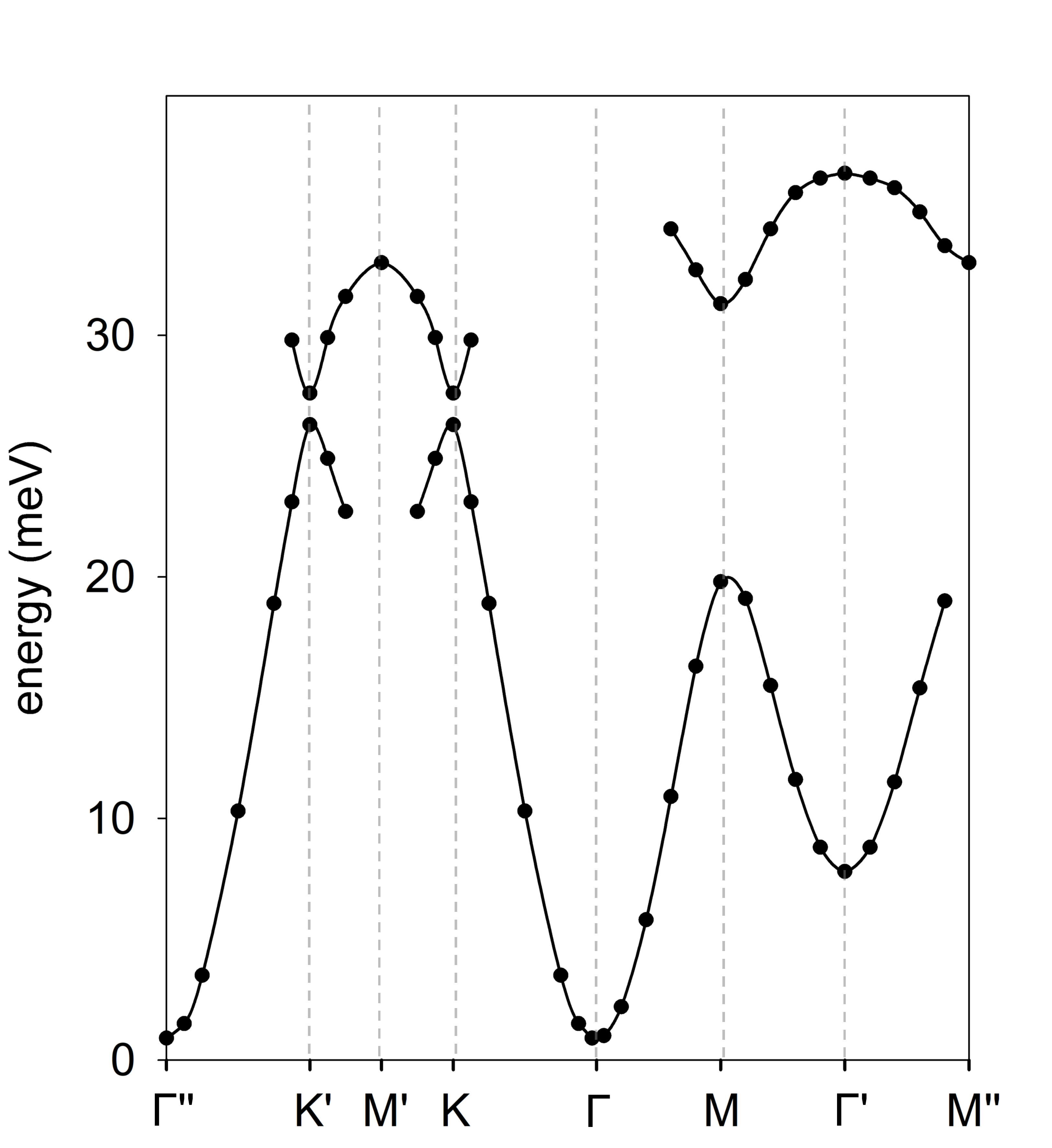}
    \caption{Magnon dispersion along the  high-symmetry directions of  the (hexagonal) Brillouin zone computed using TDDFT including SOC in the rhombohedral ferromagnetic CrI$_3$. Dots correspond to magnon energies computed explicitly using the \texttt{turboMagnon} code, while lines are the result of the interpolation.}
    \label{fig:SMpoints}
\end{figure}

We obtained the spin-wave dispersion by solving the ALDA-TDDFT equations for a finite set of wave vectors along
the 2D path
$\mathrm{\Gamma}^{\prime\prime}\!\!-\mathrm{K}^\prime\!\!-\mathrm{M}^\prime\!\!-\mathrm{K}\!\!-\mathrm{\Gamma}\!\!-\mathrm{M}\!
\!-\mathrm{\Gamma}^\prime\!\!-\mathrm{M}^{\prime\prime}$, 
and then interpolating these results on a finer grid to get a clearer picture. 
For completeness, we report
in Fig.~\ref{fig:SMpoints} the points and energies that we computed directly with the \texttt{turboMagnon} code. We chose the path for the dispersion to sample the high-symmetry directions
and points of the hexagonal lattice; thus, we have labeled them according to their position in the hexagonal Brillouin zone.
The labels refer to the high-symmetry points of the hexagonal reciprocal space; we also
report in Table~\ref{tab:SMpath} their components in terms of the rhombohedral reciprocal lattice vectors.

\begin{table}[ht]
    \centering
    \begin{tabular}{c@{~~~}c@{~~~~~}c@{~~~~~}c}
    \hline
    \hline
         \rule{0pt}{1.0em}Label & \multicolumn{3}{c}{Components}\\
         \hline
         \rule{0pt}{1.01em}$\mathrm{\Gamma}^{\prime\prime}$ & $0$ & $1$ &$-1$ \\
         $\mathrm{K}^\prime$      & $0$   & $2/3$ &$-2/3$\\
         $\mathrm{M}^\prime$      & $0$   & $1/2$ &$-1/2$\\
         $\mathrm{K}$             & $0$   & $1/3$ &$-1/3$\\
         $\mathrm{\Gamma}$        & $0$   & $0$   &$0$   \\
         $\mathrm{M}$             & $1/6$ & $1/6$ &$-1/3$\\
         $\mathrm{\Gamma}^\prime$ & $1/3$ & $1/3$ &$-2/3$\\ 
         $\mathrm{M}^{\prime\prime}$ & $1/2$ & $1/2$ & $-1$\\
    \hline
    \hline
    \end{tabular}
    \caption{The rhombohedral fractional coordinates of the hexagonal high-symmetry points used to build the path for the magnon dispersion shown in Fig.~\ref{fig:SMpoints}.}
    \label{tab:SMpath}
\end{table}

\section{Determination of the band- crossing/edge points near K}
\label{sec:min}

\begin{figure}[h]
    \hbox{
        \vbox{\hsize=0.45\textwidth
        \large{\centering (a) no SOC} \\[3pt]
        \includegraphics[width=0.9\hsize]{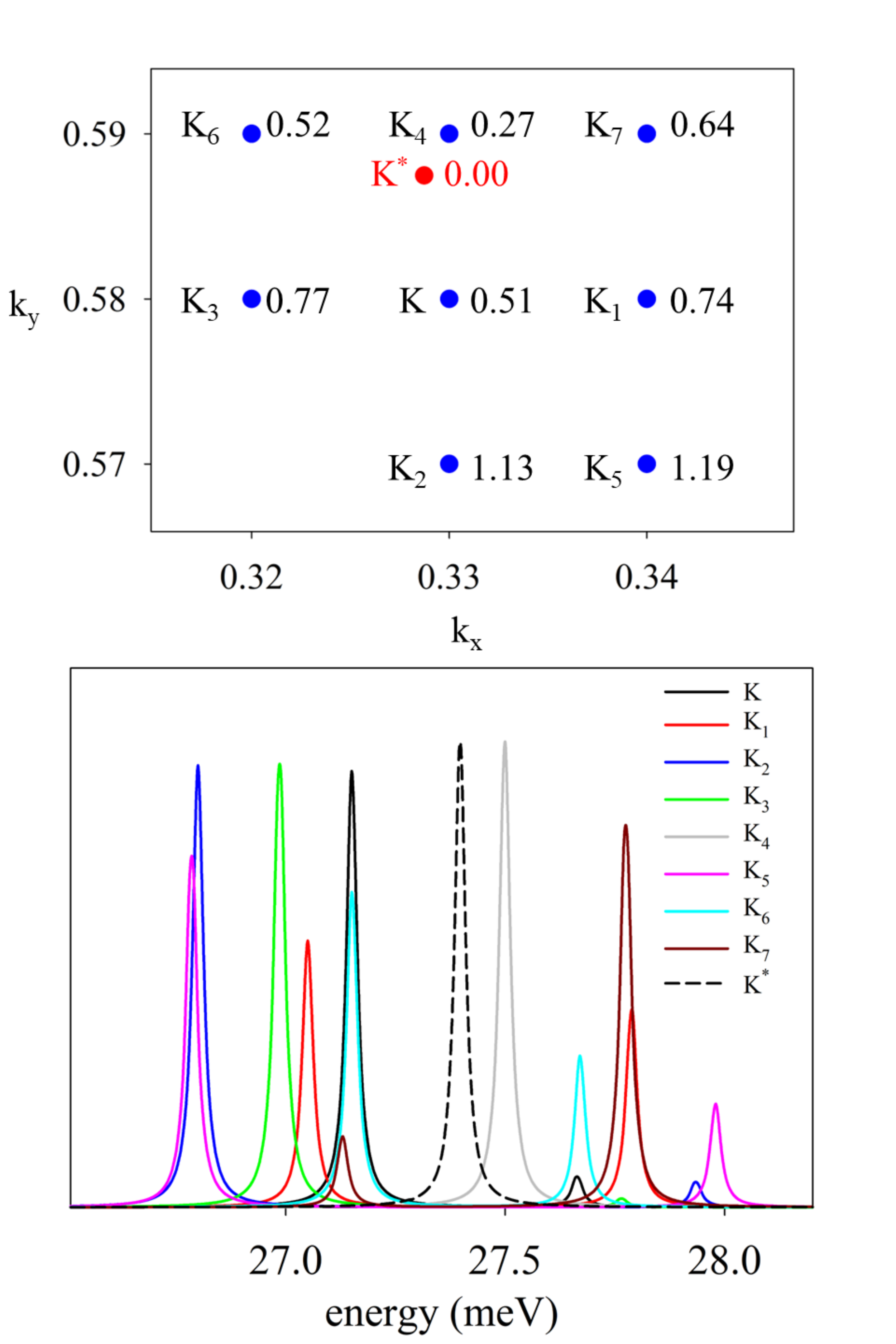}
        \vskip -10 pt
        \begin{align*}
            K^*&=(0.332,0.585) \frac{2\pi}{a} \\
            E_g&= 0
        \end{align*}
        }
        \vbox{\hsize=0.45\textwidth
        \large{\centering (b) SOC} \\[3pt]
        \includegraphics[width=0.9\hsize]{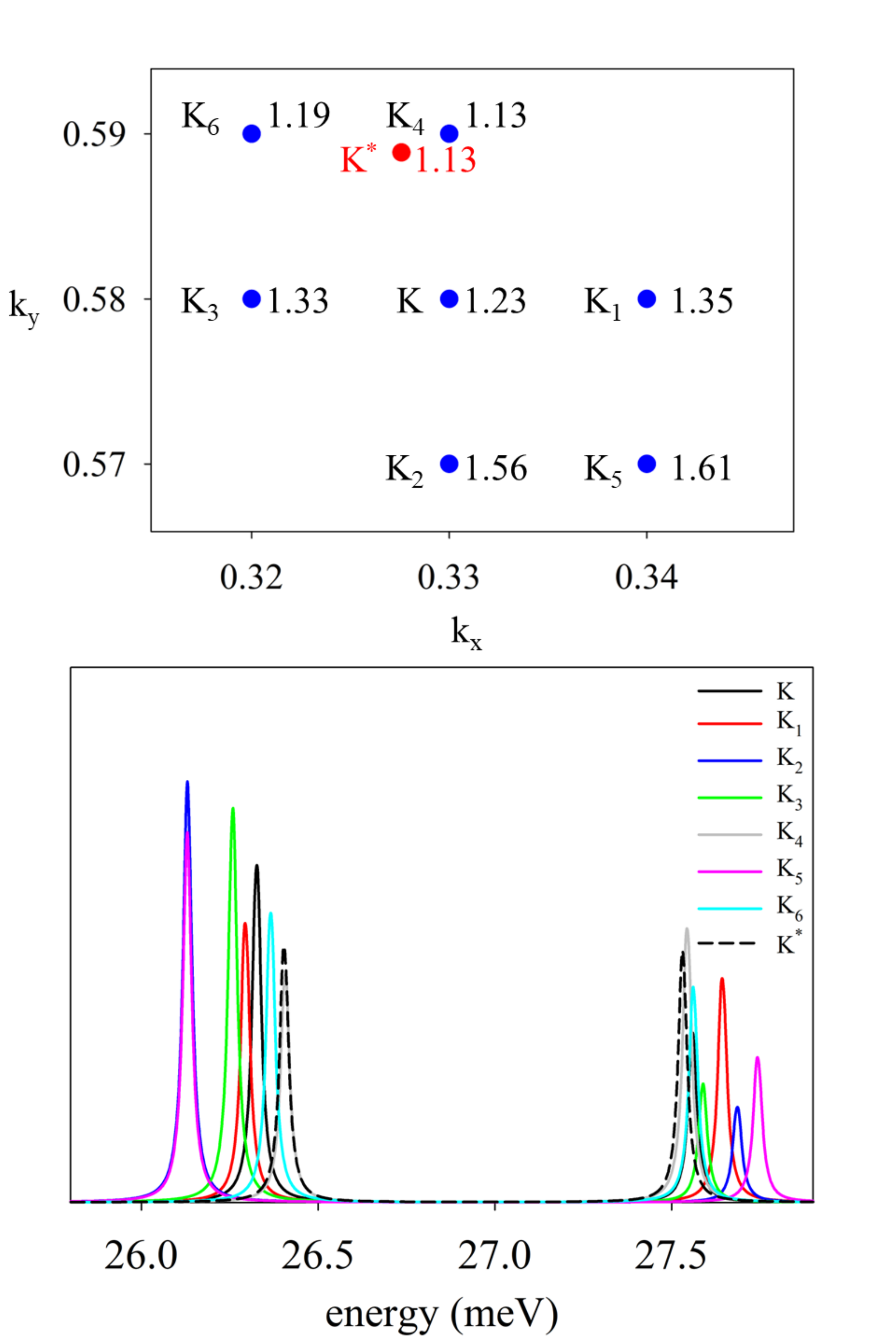}
        \vskip -10 pt
        \begin{align*}
            K^*&=(0.331, 0.586) \frac{2\pi}{a} \\
            E_g&= 1.13~\mathrm{meV}
        \end{align*}
         }
    }
    \caption{
        Upper panels: scan of the magnon spectrum of bulk CrI$_3$ in the proximity of the K point $\bm{k}_\mathbf{K} = (\frac{1}{3}, \frac{1}{\sqrt{3}}) \frac{2\pi}{a}$ in the 2D BZ, (a) neglecting SOC  and (b) accounting for them. The wavevector units are $2\pi/a$, $a=6.87~\mathrm{\AA}$ being the lattice constant of the 2D honeycomb lattice. The value of the computed gap is reported next to each point in meV. The lower panels report the computed spectra.
	} \label{fig:min}
\end{figure}

The location of the band edge/crossing near the K point in bulk CrI$_3$ and the magnitude of the gap have been determined by sampling the TDDFT magnon dispersion in the proximity of the K point, as illustrated in Fig. \ref{fig:min}, and fitted to the equation:
\begin{equation}
        \omega(\bm{k}) = \pm \sqrt{\alpha + (\bm{k} -\bm{k}_{\mathrm{K}^*})\cdot \bm{Q} \cdot(\bm{k}-\bm{k}_{\mathrm{K}^*})},
\end{equation}
where $\bm{k}$ is a point in the 2D BZ, $\bm{Q}$ is a symmetric $2\times 2$ tensor, $\bm{k}_{\mathrm{K}^*}$ the position of the band edge/crossing, and $\alpha$ a positive constant. $\mathrm{K}^*$ is a Dirac point when $\alpha=0$, a band edge otherwise.

In the the spin-orbit case, where an actual gap is present between the two branches, we have also left the freedom to fit a different  $\mathrm{\alpha}$ parameter  and $\bm{Q}$ matrix for each branch, so allowing for different $\bm{k}$-displacement between the upper and lower branch. With this fitting procedure, we obtained that the top and the bottom of the two branches have been
respectively displaced to
$K^*_{\mathit{acoustic}} = (0.3707,0.6541,0.0000)$
and 
$K^*_{\mathit{optical}} = (0.3719, 0.6570, 0.0000)$
which compares very well with our unique conoid assumption that yields a displaced extremal point at 
$K^*_{\mathit{conoid}} = (0.3714, 0.6563, 0.0000)$.
Hence, the two fitting procedures are consistent and yield an equal estimate for the value of the gap at $K^*$ and magnon band edges shifts.

\end{document}